\documentclass[a4paper]{article}
\usepackage{authblk}
\usepackage{natbib}
\usepackage[]{algorithm2e}
\usepackage{verbatim}

\usepackage[english]{babel}
\usepackage[utf8x]{inputenc}
\usepackage[T1]{fontenc}

\usepackage[a4paper,top=3cm,bottom=2cm,left=3cm,right=3cm,marginparwidth=1.75cm]{geometry}

\usepackage{amsmath}
\usepackage{rotating, graphicx}
\usepackage{caption}
\usepackage{subcaption}
\usepackage[colorinlistoftodos]{todonotes}
\usepackage[colorlinks=true, allcolors=blue]{hyperref}
\usepackage{multirow}
\usepackage[toc,page]{appendix}

\usepackage{color}
\usepackage{soul}
\usepackage{array}
\usepackage{amssymb}

\usepackage{csquotes}
\usepackage{breakcites}

\begin{document}

\title{Exploring and comparing temporal clustering methods}

\author[1,2]{Jordan Cambe}
\author[1,2]{Sebastian Grauwin}
\author[1]{Patrick Flandrin}
\author[1,2]{Pablo Jensen\footnote{Corresponding author}}

\affil[1]{Univ Lyon, ENS de Lyon, UCB Lyon 1, CNRS, Laboratoire de Physique, F-69342 Lyon, France}
\affil[2]{Institut Rh\^{o}nalpin des Systemes Complexes, IXXI, F-69342 Lyon, France}

\maketitle

\begin{abstract}
Description of temporal networks and detection of dynamic communities have been hot topics of research for the last decade. However, no consensual answers to these challenges have been found due to the complexity of the task. Static communities are not well defined objects, and adding a temporal dimension renders the description even more difficult. In this article, we propose a coherent temporal clustering method: the Best Combination of Local Communities (BCLC). Our method aims at finding a good balance between two conflicting objectives : closely following the short time evolution by finding optimal partitions at each time step and temporal smoothness, which privileges historical continuity. We test our algorithm on two bibliographic data sets by comparing their mesoscale dynamic description to those derived from a (static) simple clustering algorithm applied over the whole data set. We show that our clustering algorithm can reveal more complex dynamics than the simple approach and reach a good agreement with expert's knowledge.

\end{abstract}


\section{Introduction}
Networks are a convenient way to represent real-world complex systems, such as social interactions \cite{newman_2003},\cite{onnela_2007}, metabolic interactions \cite{boccaletti_2006}, the Internet/world wide web \cite{vespignani_2001}, transportation systems \cite{barrat_2006,barthelemy_2011}, etc. For several systems it is interesting to find and describe areas of the network which are more densely connected, i.e. the communities of the network. In 20 years of complex networks history extensive work was conducted on community detection in static - non evolving - networks, see \cite{newman_2006,blondel_2008,fortunato_2007} and the review \cite{fortunato_2016} for an overview on community detection in static graphs. 

However, many networks have a temporal dimension and need a dynamic mesoscopic description at risk of non-negligible information losses if studied as static networks. Therefore the description of large temporal graphs has been a hot topic of research for the last decade, see the excellent reviews \cite{HolmeSaramaki_2012} and \cite{Holme_2015} for a complete description of temporal networks. Most recently the detection of dynamic communities, that is communities on temporal networks, has become one of the main interests in network science, as temporal networks require to adapt the methods of static community detection. So far no consensual method was found and around 60 methods have been proposed to try to detect dynamic communities evolving with temporal networks. A total of four published reviews try to classify and summarize them \cite{Aynaud_2013}, \cite{Hartmann_2016}, \cite{Masuda_2016} and \cite{Rossetti_2018}. 

In \cite{Rossetti_2018}, these methods are classified into 3 main categories: (a) \emph{instant optimal}, (b) \emph{temporal trade-off} and (c) \emph{cross-time}. These methods aim to detect clusters at different times t, i.e. for many snapshots of the temporal network. As these clusters are only dependent on the state of the network at time t, it is then necessary to match the communities at different t with some similarity measures, e.g. Jaccard based \cite{morini_2017, lorenz_2017, greene_2010}, core-node \cite{wang_2008}. Methods in category (b) define clusters at t depending on current and past states of the network. Clusters are incrementally temporally smooth. However such methods are subject to drift as clusters are added up to each other locally. There is no compromise between temporal smoothness and 'optimal' partition at time $t$, see for example \cite{rossetti_2017,guo_2014,gorke_2010,gorke_2013}. Finally, in category (c) clusters at $t$ depend on both past and future states of the network, see \cite{duan_2009,mucha_2010,matias_2016,ghasemian_2016}. Clusters are completely temporally smooth and not subject to drift, but they do not respect causality as communities at $t$ are determined using network's information at $t+n$, i.e. communities at time $t$ can change depending on what comes next, which makes these methods inappropriate for use on-the-fly.

In this article, we present a new tool to achieve a mesoscopic description of dynamic networks, which tries to find a good compromise between 'global' and 'local' methods. We apply our method on two data sets of scientific articles, to show how it can describe the emergence and evolution of scientific disciplines. The main difficulties for meta-community detection methods are twofold: Finding the right temporal smoothing and quantifying the `stability' of communities. It is difficult to distinguish if changes between snapshots are due to structural evolution of the community or algorithm instability, as static community detection methods used at each time $t$ can find different communities for a same topology (see \cite{Rossetti_2018} for a complete description of pros and cons of each clustering category). Here, we propose an algorithm which aims to find a good balance between temporal inertia (smoothness) and `optimal' partition at any given time $t$. We compare this method to the most basic approach, which optimizes the modularity of the aggregated network using the Louvain algorithm \cite{blondel_2008}. The latter can be assimilated to a category (c) method in \cite{Rossetti_2018}. We then describe the methods to analyze differences between partitions: mutual information (MI) measures and bipartite network (BN) representations. We show that MI based measures are interesting but give a limited amount of information on how different two partitions are, whereas bipartite network representation allows to see how streams split between partitions. We used the methods on two bibliographic data sets: (1) the scientific publications of a scientific institution, ENS Lyon and (2) publications related to the emergence of a new mathematical tool, the 'wavelets'. We show that the global approach represents a good approximation when the dynamics is simple, i.e. when there are mainly parallel streams without much interaction, as in the ENS Lyon case. However, when the dynamics is more complex (and interesting), when many communities are born, die, split or merge, one needs a more sophisticated approach, and we show that our algorithm performs well compared to an expert-based dynamics.

\section{Methods} 
We start by presenting the two building blocks used in the algorithms we want to compare: how we define and partition a Bibliographic Coupling (BC) network and how we match clusters from successive time periods to create `streams' (temporal meta-clusters). We then introduce the \emph{BiblioMaps} platform used for visualizing dynamic communities and finally we present two standard methods to compare partitions derived from the different methods : Normalized Mutual Information and Bipartite Networks.

\subsection{Bibliographic Coupling partitioning}
Given a set of publications on a given period, a Bibliographic Coupling (BC) network can be defined based on the relative overlap between the references of each pair of publications. More specifically, we compute Kessler's similarities $\omega_{ij}=R_{ij}/\sqrt{R_{i}R_{j}}$, where $R_{ij}$ is the number of shared references between publications $i$ and $j$ and $R_{i}$ is the number of references of publication $i$. In the BC network, each publication corresponds to a node and two publications $i$ and $j$ share a link of weight $\omega_{ij}$. If they don't share any reference, they are not linked ($\omega_{ij}=0$); if they have an identical set of references, their connexion has a maximal weight ($\omega_{ij}=1$). Here, we consider that the link between two publications is only meaningful if they share at least two references and we impose $\omega_{ij}=0$ if they share only one reference.\\

We use weighted links to reinforce the dense (in terms of links per publication) regions of the BC networks. This reinforcement facilitates the partition of the network into meaningful groups of cohesive publications, or communities. We measure the quality of the partition with the \emph{modularity} Q (eq. \ref{eq1}), a quantity that roughly compares the weight of the edges inside the communities to the expected weight of these edges if the network were randomly produced:

\begin{equation}
Q = \frac{1}{2\Omega} \sum_{i,j} \left[ \omega_{i,j} - \frac{\omega_i \omega_j}{2\Omega} \right]\delta\left(c_i, c_j \right),
\label{eq1}
\end{equation}

\noindent where $\omega_i = \sum_j \omega_{ij}$ is the sum of the weights of the edges linked to node i, $c_i$ and $c_j$ are the communities containing respectively nodes $i$ and $j$, $\delta$ is the Kronecker function ($\delta (u,v)$ is 1 if $u = v$ and 0 otherwise) and $\Omega = \frac{1}{2}\sum_{ij} \omega_{ij}$ is the total weight of edges. We compute the graph partition using the efficient heuristic algorithm presented in \cite{blondel_2008}.

\subsection{Matching communities from successive time periods}
Given the sets of communities $\{C_1^t, ...C_{k_t}^t \}$ in each time window $t$, the problem at hand is to identify a set of relevant historical communities, or \emph{streams}, that correspond to a chain of communities from successive time periods (at most one per period). In order to decide which community of a given period should be added to a chain of communities from previous periods, we need to use some measure of the similarity between communities from different time periods. A standard measure is the Jaccard index, which computes the proportion of shared nodes between clusters of successive and overlapping periods (see e.g. \cite{gingras_2016,morini_2017}). One drawback of this measure is the need to use overlapping periods which implies that there is no bijection between the publications and the streams (a given publication can belong to several streams).\\

Here, we take advantage of the fact that links can be computed between nodes from different time periods (publications from different periods can have common references). We could for example define a similarity measure between two clusters $C_a$ and $C_b$ from different periods either by the total sum of the links between pairs of publications from these clusters $\Omega_{a,b} = \sum_{i \in C_a, j \in C_b} \omega_{i,j}$ or by a normalized version of this sum  $\omega_{a,b}=\Omega_{a,b} /|C_a||C_b|$, which is comprised between 0 and 1. While these two measures may appear quite intuitive, each of them has some drawbacks as well: using $\Omega_{a,b}$ may bias the construction of the streams by linking two `large' (in terms of publications) but rather dissimilar (in terms of shared references) clusters. On the opposite, using $\omega_{a,b}$ may create some biases by linking two very similar clusters of very different sizes rather than the two clusters that have the second-best similarity and have similar sizes. To be coherent with our construction of clusters (maximizing the modularities within each time period), we propose here to use a modularity-based concept to match clusters from successive time periods. The similarity measure we use is thus $\delta Q = \Omega_{a,b} - \Omega_{a}\Omega_{b}/2\Omega_{A,B}$ which corresponds to an increase in the modularity of the BC network built from the two periods $A$ and $B$ when, starting from partitions defined on each period, clusters $a$ and $b$ are merged.\\

{\bf Matching Algorithm}\\
\begin{algorithm}[H]
  Only compare pairs of communities $(a,b)$ with a minimum similarity $\omega_{a,b}> \Theta=10^{-6}$.\\
  Define the best match of each cluster by the one maximizing $\delta Q$.\\
 \For{each temporal window}{
  Define the {\bf predecessor} of each cluster as its best match from the previous time period.\\
  Define the {\bf successor} of each cluster as its best match from the next time period.
  }
 Two clusters are said to be \emph{paired} if they are predecessors / successors of each other.\\
 If a cluster is not the successor of its predecessor, we have a \emph{split}.\\
 If a cluster is not the predecessor of its successor, we have a \emph{merge}.\\
 Streams are defined as chains of paired clusters.\\ 
\end{algorithm}

\begin{figure}
\centering
\resizebox*{12cm}{!}{\includegraphics{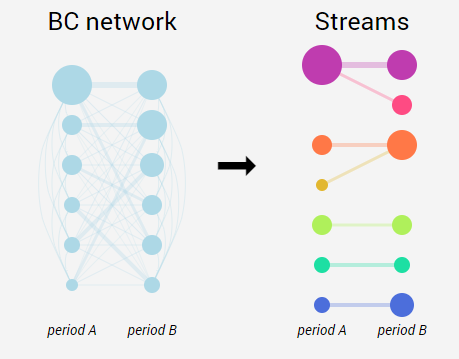}}
\caption[Matching clusters from successive time periods.]{Matching clusters from successive time periods. We start with the BC network built from clusters of publications detected independently in periods A and B. For each cluster $a$ of period A and cluster $b$ of period B, we compute the modularity change $\delta Q_{ab}$ obtained by grouping these two clusters in the 2-periods BC network. The `successor' of cluster $a$ is defined as the cluster $b$ from period B maximizing this quantity and the `predecessor' of cluster $b$ is defined as the cluster $a$ from period B maximizing this quantity. In the stream visualisation, `paired' clusters (successor / predecessor of each other) are represented on the same y-position, and we only show the BC links between successors or predecessors, which highlight dynamical events such as merge and splits. In these figures, the size of the nodes are proportional to the number of publications in the corresponding clusters and the thickness of the links represent the average weights of the BC links between publications from two clusters.}
\label{fig:matching}
\end{figure}

An illustration of this algorithm is given in Figure \ref{fig:matching}.

\subsection{Different algorithms used to define historical streams}
We compare the results of two types of algorithms which build historical communities, or \emph{streams}, starting from publications data sets. The first method is `global', as it considers the whole data set to compute the communities. The second is `local', as it starts from successive windows of $\Delta T$ years and starts by building a mesoscopic description adapted to that specific window. Hereafter, we present the results from the two main variants and refer the reader to the Annex \ref{annex1} for more detailed results.\\

\subsubsection{Global Algorithm (GA)}
The Global Algorithm builds a $global$ BC network by taking into account all the publications in the data set. Streams are defined as time evolution of these (static) communities maximizing the global modularity. Since we are working in a single (large) time period, this approach does not yield any dynamical events such as splitting / merging of communities, but it provides a simple reference. 

\subsubsection{Best-Combination Local Communities (BCLC)}
This $local$ algorithm starts by running, for each time period, $N$ independent runs (we used $N=100$) of the Louvain algorithm. Because of the noise inherent to the Louvain algorithm, the best modularity partitions in each time period are not necessarily the ones that best match each other across successive time periods. We thus optimize the inter-period combination by the following algorithm: \\

{\bf BCLC Algorithm}\\
\begin{algorithm}[H]
Compute the Bibliographic Coupling Graph \;
Split the data set into temporal windows $\Delta t$ \;
\For{each of the $N=100$ partitions of the first period and each of the $N=100$ partitions of the second period}{
 Run the matching algorithm to define the 2-periods streams\;
}
Among the $N*N$ defined streams, select the ones maximizing the modularity of the BC network on the first 2 periods. \;
Define the `best combination' partitions of the first 2 periods as those corresponding to those streams\;

\For{each pair of successive temporal windows $A$ and $B$, starting from the second one}{
   \For{each of the $N=100$ partitions of the period $B$}{
      Run the matching algorithm between these partitions and the `best combination' partition of period $A$ (known from a previous step) to define 2-periods streams\;}
 Among the $N$ defined streams, select the ones maximizing the modularity of the BC network on periods $A$ and $B$\; 
 Define the `best combination' partition of period $B$ as the one corresponding to those streams
}
  \
\end{algorithm}

Note that maximizing a global indicator over the $T$ periods with $N$ runs would take too long as there would be $N^T$ possibilities to explore. For this reason, we choose the best combinations between the first two periods ($N^2$ checks) and then one period at a time ($N(T-2)$ checks).\\

\begin{figure}
\centering
\resizebox*{12cm}{!}{\includegraphics{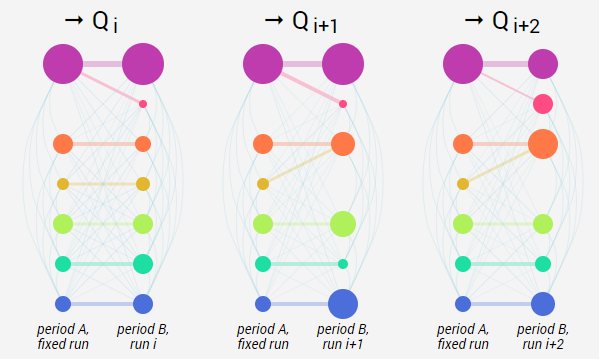}}
\caption[Choosing the best combination.]{Choosing the best combination. Given a set of clusters from period A, we perform the matching algorithm between these clusters and N=100 sets of clusters from period B obtained by independent runs of the Louvain algorithm. In each case, we compute the modularity $Q$ obtained by grouping paired clusters in a single clusters in the 2-periods BC network. The `best combination' is defined as the set of clusters from period B maximizing this quantity.}
\label{fig:bestcombinaison}
\end{figure}

This algorithm returns temporal streams we call \emph{BCLC-streams}. These streams still maximize the modularity at each time $t$ while using some cross-time information to improve the global modularity. Figure \ref{fig:bestcombinaison} is an illustration of different runs of this algorithm. Choosing the value of the period $T$ is a trade-off. It needs to be long enough so that communities within each period have enough articles to be meaningful and limit clustering variability. But it also needs to be small enough to follow scientific dynamics. For example, the mean cited half-life of scientific articles is close to 7 years \cite{half_life}. After trying different periods, we chose, for our databases, a period {$T = 5$} years.\\

\subsection{BiblioTools / BiblioMaps}\label{sec:bibliotools}
All the data sets were extracted from the ISI Web of Knowledge Core Collection database\footnote{\textit{http://apps.isiknowledge.com/}}. The bibliographic records were parsed and analyzed using Bibliotools, a Python-based open-source software and the historical streams figures were generated using the web-based visualisation platform BiblioMaps \cite{Grauwin2011,lund2017carte,grauwin2018bibliomaps}. Bibliotools and its extension BiblioMaps were developed by one of us and are available online, as all the data analysis presented in this paper\footnote{\textit{http://www.sebastian-grauwin.com/bibliomaps/}}. 

\subsection{Normalized Mutual Information}
The mutual information (MI) is a widely used measure for comparing community detection algorithms. It is defined as a measure of the statistical independence between two random variables (see eq. \ref{eq2}). In other words, if $H(P_{X})$ is the entropy associated with partition $X$ and $H(P_{Y})$ is the entropy associated with partition $Y$ (the entropy is a measure of how partitioned is our network, the more communities - here temporal streams - the higher the entropy), then $MI(P_{X},P_{Y})$ represents the overlap of the two partitions. In layman's terms, it tells us how much we know about the partition $P_{X}$ when the partition $P_{Y}$ is given. You may refer to  \cite{Wagner_2007,Kvalseth_2017} for a deeper description on mutual information. In particular, note that the mutual information is a symmetrical measure

\begin{equation}
MI(P_{X},P_{Y}) = H(P_{Y}) - H(P_{Y}\vert P_{X}) = MI(P_{Y},P_{X})
\label{eq2}
\end{equation}
\ 

The MI is defined on $[0,+\infty]$, therefore it is difficult to make sense of it without an upper-bound. There exists different ways to normalize the mutual information. The idea is to take into account the entropies of the partitions we consider to gauge the proportion of mutual information between the partitions. Normalizing by the entropy of one of the partition, e.g. $H(P_{X})$ (see eq. \ref{eq3}) measures how much of the partition $P_{X}$ is included in the partition $P_{Y}$. We call this normalized mutual information $NMI_{X}$. If it reaches its maximum value 1, it means that it is possible to retrieve all the information (the partition) of $P_{X}$ from the partition $P_{Y}$. However this measure does not take into account the size of the other partition, $P_{Y}$. A partition $P_{Y}$ where each node would be its own community would make $NMI_{X}$ equals to 1 even though both partitions are very different. This measure then needs to be combined with at least another NMI which takes into account the relative size of both partitions (see eq. \ref{eq4}). Here the mutual information is normalized by $\sqrt{H(P_{X})*H(P_{Y})}$, which shows how much of the two entropies overlap on a scale between 0 and 1. This expresses how similar the partitions are. It is equal to 1 when the partitions are the same. Moreover, this last NMI is symmetrical, so it takes into account both retrieval of $P_{X}$ from $P_{Y}$ and retrieval of $P_{Y}$ from $P_{X}$.

\begin{equation}
NMI_{X}(P_{X},P_{Y}) = \frac{MI(P_{X},P_{Y})}{H(P_{X})}
\label{eq3}
\end{equation}

\begin{equation}
NMI(P_{X},P_{Y}) = \frac{MI(P_{X},P_{Y})}{\sqrt{H(P_{X})*H(P_{Y})}} = NMI(P_{Y},P_{X})
\label{eq4}
\end{equation}
\ 
While Mutual information based measures give a value of similarity between two partitions, it is not straightforward to analyze. For example, it does not allow to track where the (dis)similarities come from. To allow in depth comparison, we represent pairs of partitions as bipartite networks.

\subsection{Bipartite Network of streams}

To track and quantify differences between partitions $X$ and $Y$, we compute a bipartite network where the $n_{X}^i \in N_{X}$ are the first kind of nodes. They represent the streams $s_{X}^i \in P_{X}$ (hence $\vert N_{X}\vert = \vert P_{X}\vert$). It follows that the second kind of nodes $n_{Y}^j \in N_{Y}$ represent the streams $s_{Y}^j \in P_{Y}$. A weighted directed edge is drawn between $n_{X}^i$ and $n_{Y}^j$ only if their corresponding streams $s_{GA}^i$ and $s_{BCLC}^j$ share articles. For a given pair of nodes ($n_{X}^i$,$n_{Y}^j$) the weights of the two edges between them (one in each direction) are defined in eq.\ref{eq5}. We quantify differences between streams of two partitions from this graph, quantities are given in table \ref{tab:flows}.

\begin{equation}
  \left\{
      \begin{aligned}
			w_{n_{X}^i\rightarrow n_{Y}^j} = \frac{\vert s_{X}^i \cap s_{Y}^j\vert}{\vert s_{X}^i\vert}\\
			w_{n_{Y}^j\rightarrow n_{X}^i} = \frac{\vert s_{Y}^j \cap s_{X}^i\vert}{\vert s_{Y}^j\vert}
      \end{aligned}
    \right.
\label{eq5}
\end{equation}

\section{Data sets}\label{sec:dyncom_data set}
In this section, we present the specificity of each data set and the motivations for studying them. Key statistical data are summarized in table \ref{tab:data sets}.

\begin{table}
\centering
\resizebox{\textwidth}{!}{
\begin{tabular}{rrrrrrrrrr}
\hline
data set & Type & Period & N & $N_{BC}$ & $\rho_{links}$ & $< d >$ & $< w >$ & $Q_{GA}$\\
\hline
Wavelets & Thematic & 1963-2012 & 6,582 & 5,568 & 0.0065 & 35.98 & 0.000719 & 0.677\\
ENS-Lyon & Institution & 1988-2017 & 16,679 & 14,389 & 0.0019 & 27.04 & 0.000175 & 0.919\\
\hline
\end{tabular}}
\caption[Statistics on data sets investigated]{Statistics on data sets investigated in the paper. \emph{Type} is the type of organization data come from. \emph{Period} is the period over which spans the data set. \emph{N} is the number of publications in the data set. \emph{$N_{BC}$} is the number of articles in the BC table. $\rho_{links}$ is the density of links in the BC network. $< d > = (N_{BC}-1)*\rho_{links}$, it indicates the average number of publications a given publication shares references with. $< w >$ indicates the average link weight. \emph{Q} is the modularity of the network using global partitioning (GA).}
\label{tab:data sets}
\end{table}

\subsection{ENS-Lyon Publications data set}
The ENS-Lyon Publications data set contains all publications produced by researchers affiliated to the \emph{\'Ecole Normale Sup\'erieure de Lyon} in natural science fields. It spans the 1988-2017 period and contains 16,679 publications. As for many scientific institutions, its publication records is highly structured by disciplinary academic departments. Here, we compare our temporal clustering methods to a partition that clusters articles according to their authors' laboratories (reference partition, $P_{REF}$).

\begin{sidewaysfigure}
\begin{subfigure}[b]{\textwidth}
\centering
\resizebox*{20cm}{!}{\includegraphics{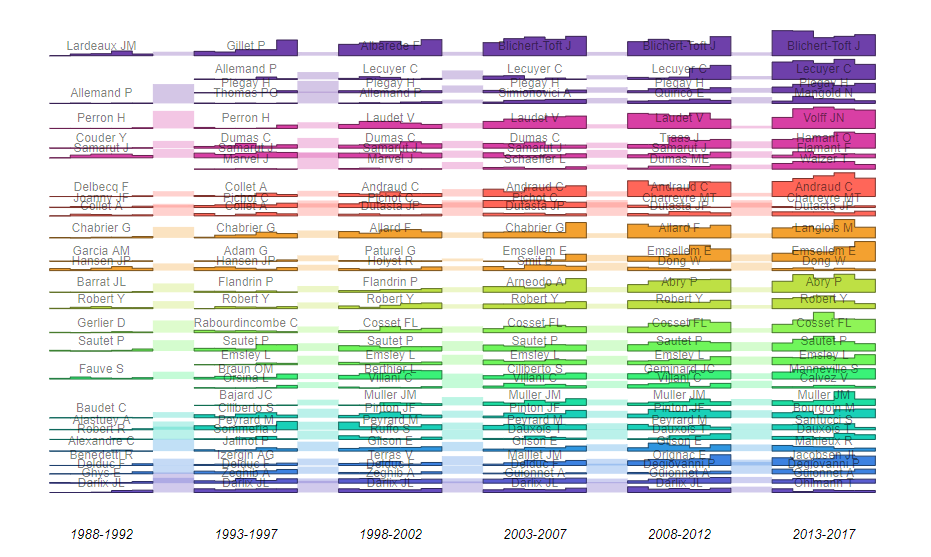}}
\caption[Historical streams computed from the ENS Lyon natural sciences publications]{Historical streams computed from the ENS Lyon natural sciences publications. Streams were determined using the global algorithm (GA).}
\label{fig:streams_ensl_a}
\end{subfigure}
\end{sidewaysfigure}

\begin{sidewaysfigure}\ContinuedFloat
\begin{subfigure}[b]{\textwidth}
\centering
\resizebox*{20cm}{!}{\includegraphics{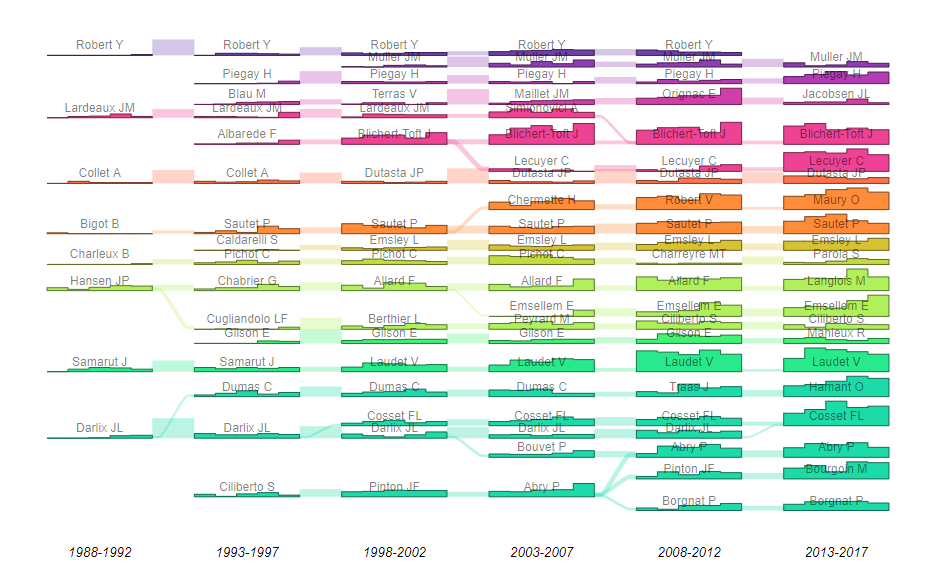}}
\caption[Historical streams computed from the ENS Lyon natural sciences publications]{Historical streams computed from the ENS Lyon natural sciences publications. Streams were computed using our local method (BCLC).}
\label{fig:streams_ensl_b}
\end{subfigure}
\caption[Historical streams computed from the ENS Lyon natural sciences publications]{Labels on each stream correspond to the most frequent author name in that stream during a given period. Streams with the same color have close research topics (here the proximity of streams to each other is computed from the weight of BC network links between clusters of a same period). Bar height is proportional to the number of publications in a given year. Links between streams show the streams that are preceding/following each other.}
\label{fig:streams_ensl}
\end{sidewaysfigure}

\subsection{Wavelets Publications data set}
The Wavelets Publications data set contains all publications related to wavelets and spans from 1910 to 2012 (however the period before 1960 contains only a few publications). This data set contains 6,582 publications, corresponding to all the publications of a list of 83 key actors in the field of wavelets selected by expert advice and bibliographic searches (for more details, see \cite{morini_2017}). The study of this data set represents a difficult task because it emerged from the collaboration of several research fields, constituted by many entangled subfields. Based on the knowledge of one of the authors (PF), a field's expert, we built manually a temporal partition drawing the history of wavelets. We refer to this partition as $P_{REF}$ and compare our automatically generated partitions to this partition of reference. We acknowledge that this partition is not an absolute ground truth as it relies on the subjectivity of an expert. However, we assume that this reference gives a reasonable picture of the field’s evolution.

\begin{sidewaysfigure}
\begin{subfigure}[b]{\textwidth}
\centering
\resizebox*{23cm}{!}{\includegraphics{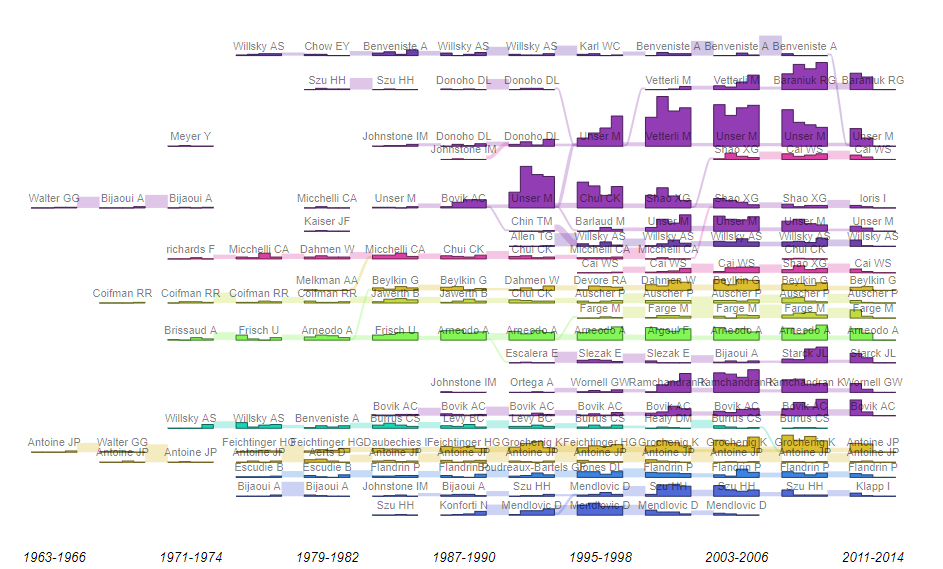}}
\caption[Historical streams computed from the wavelets field of research publications]{Historical streams computed from the wavelets field of research publications. Streams determined using the global algorithm (GA).}
\label{fig:streams_wavelets_a}
\end{subfigure}
\end{sidewaysfigure}

\begin{sidewaysfigure}\ContinuedFloat
\begin{subfigure}[b]{\textwidth}
\centering
\resizebox*{23cm}{!}{\includegraphics{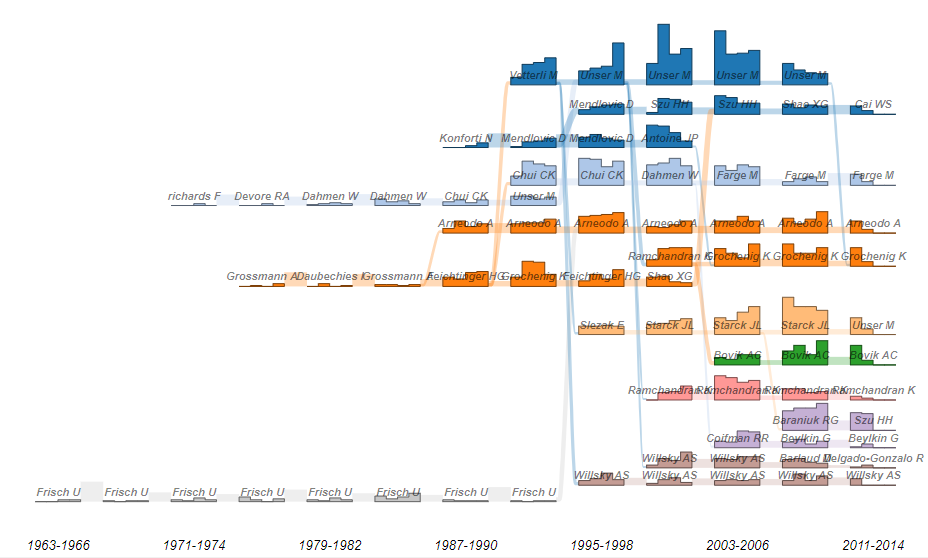}}
\caption[Historical streams computed from the wavelets field of research publications]{Historical streams computed from the wavelets field of research publications. Streams were computed using our local method (BCLC).}
\label{fig:streams_wavelets_b}
\end{subfigure}
\end{sidewaysfigure}

\begin{sidewaysfigure}\ContinuedFloat
\begin{subfigure}[b]{\textwidth}
\centering
\resizebox*{23cm}{!}{\includegraphics{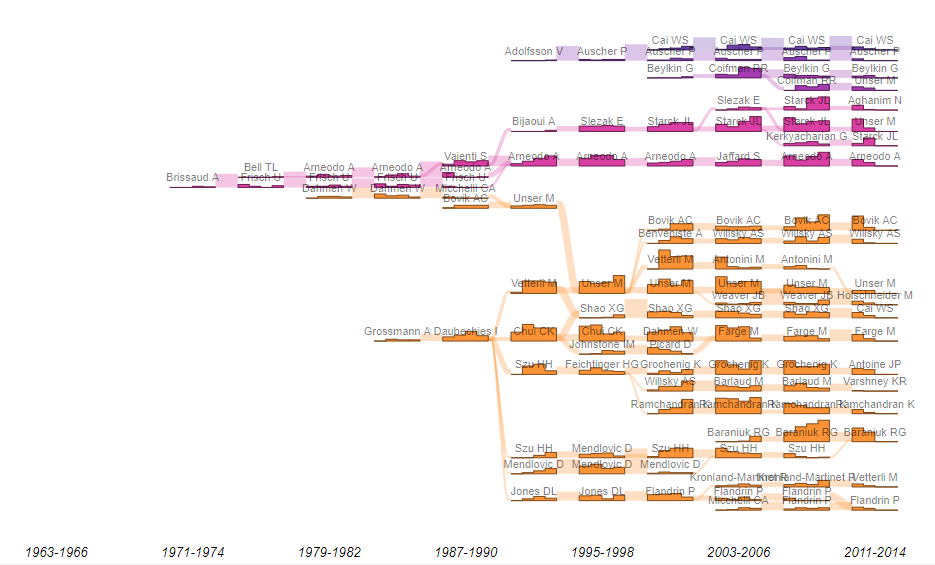}}
\caption[Historical streams computed from the wavelets field of research publications]{Historical streams computed from the wavelets field of research publications. Streams are the reference streams determined manually.}
\label{fig:streams_wavelets_c}
\end{subfigure}
\caption[Historical streams computed from the wavelets field of research publications]{Labels on each stream correspond to the most frequent author name in that stream during a given period. Streams with the same color have close research topics (here the proximity of streams to each other is computed from the weight of BC network links between clusters of a same period). Bar height is proportional to the number of publications in a given year. Links between streams show the streams that are preceding/following each other.}
\label{fig:streams_wavelets}
\end{sidewaysfigure}

\section{Results}

It is difficult to represent the richness of the information conveyed by streams in paper figures. To be able to attribute scientific meaning to each of the streams, and characterize them through their main authors, references, keywords... an interactive stream visualization is available at {\textit{ http://www.sebastian-grauwin.com/streams/BCstreams.html}}.

\subsection{General features}

As illustrated in Figures \ref{fig:streams_ensl} and \ref{fig:streams_wavelets}, the global method cannot lead to a rich dynamics description. By construction, GA streams are well separated from each other (Figure \ref{fig:streams_ensl_a}) and show only a few links in Figure \ref{fig:streams_wavelets_a}, which could be interpreted as splits or merges of subfields. On the opposite, BCLC streams lead to a more dynamical history for both data sets. There are only a few links Figure \ref{fig:streams_ensl_b}, because different streams correspond to different scientific (sub)disciplines, which are known to be only marginally connected. However, our method rightly spots teams that split to focus on different research topics (as streams `Blichert-Toft' and `Lecuyer', 6$^{th}$ and 7$^{th}$ from the top). Similarly, many splits and merges occur in Figure \ref{fig:streams_wavelets_b}. To analyze these differences, we compare the partitions for each data set using the two measures defined above : Mutual information (table \ref{tab:NMIs}) and community similarity from the bipartite network representation.

\begin{table}
\centering
\begin{tabular}{rrr}
\hline
Measures & ENS-Lyon & Wavelets \\
\hline
$\vert P_{GA}\vert$ & 57 & 27 \\
$\vert P_{BCLC}\vert$ & 97 & 36 \\
$\vert P_{REF}\vert$ & 17 & 36 \\
\hline
$H(P_{GA})$ & 3.63 & 2.87 \\
$H(P_{BCLC})$ & 4.05 & 3.04 \\
$H(P_{REF})$ & 2.37 & 3.18 \\
\hline
$MI(GA,REF)$ & 1.93 & 2.03 \\
$MI(BCLC,REF)$ & 1.93 & 2.49 \\
$MI(GA,BCLC)$ & 3.10 & 1.90 \\
\hline
$NMI_{REF}(GA,REF)$ & 0.82 & 0.64 \\
$NMI_{REF}(BCLC,REF)$ & 0.81 & 0.80 \\
$NMI(GA,BCLC)$ & 0.81 & 0.64 \\
\hline
\end{tabular}
\caption[Entropies and Mutual Information measures]{$\vert P_{X}\vert$ is the number of streams in partition $X$. $H(P_{X})$ is the entropy of partition $X$. $MI(P_{X},P_{Y})$ is the mutual information between the partitions $X$ and $Y$. $NMI_{REF}$ is the mutual information MI normalized by $H(P_{REF})$. $NMI(GA,BCLC)$ is the symmetrical normalized mutual information (normalized by $\sqrt{H(GA)*H(BCLC)}$).}
\label{tab:NMIs}
\end{table}

\subsection{Results on ENS Lyon data set}
Table \ref{tab:NMIs} shows the highly different number of streams of each partition : 57 streams for the global method, 97 for the local one and only 17 for the reference partition (the 17 laboratories of the ENS Lyon). The high values of $NMI_{REF}(GA,REF)$ (0.82) and $NMI_{REF}(BCLC,REF)$ (0.81) suggest that the extra streams in both $P_{GA}$ and $P_{BCLC}$ are mostly hierarchical subdivisions of the laboratory streams from $P_{REF}$. A partition being a subdivision of another does not result in a decrease of MI between them. The MI decreases only if communities of a partition need to be mixed to become communities of another. These results suggest that $P_{GA}$ and $P_{BCLC}$ are merely a smaller-scale division of $P_{REF}$. Similarly, the high value of $NMI(GA,BCLC)$ (0.81) suggests that $P_{BCLC}$ and $P_{GA}$ convey the same information.

The measures from Table \ref{tab:flows} confirm this analysis. $\overline{1^{st}E}(GA,REF)$ shows that streams from $P_{GA}$ share on average $86 \pm 17\%$ of their articles with a stream from $P_{REF}$ and an average of $3.37\pm 1.76$ streams from $P_{GA}$ are needed to retrieve 80\% of streams from $P_{REF}$. Similar observations can be made for $P_{BCLC}$. Moreover, $\overline{Sum_{80}}(GA,BCLC)$ shows that it takes on average two streams from $P_{BCLC}$ to reach 80\% of streams from $P_{GA}$. 

\begin{figure}
\centering
\resizebox*{12cm}{!}{\includegraphics{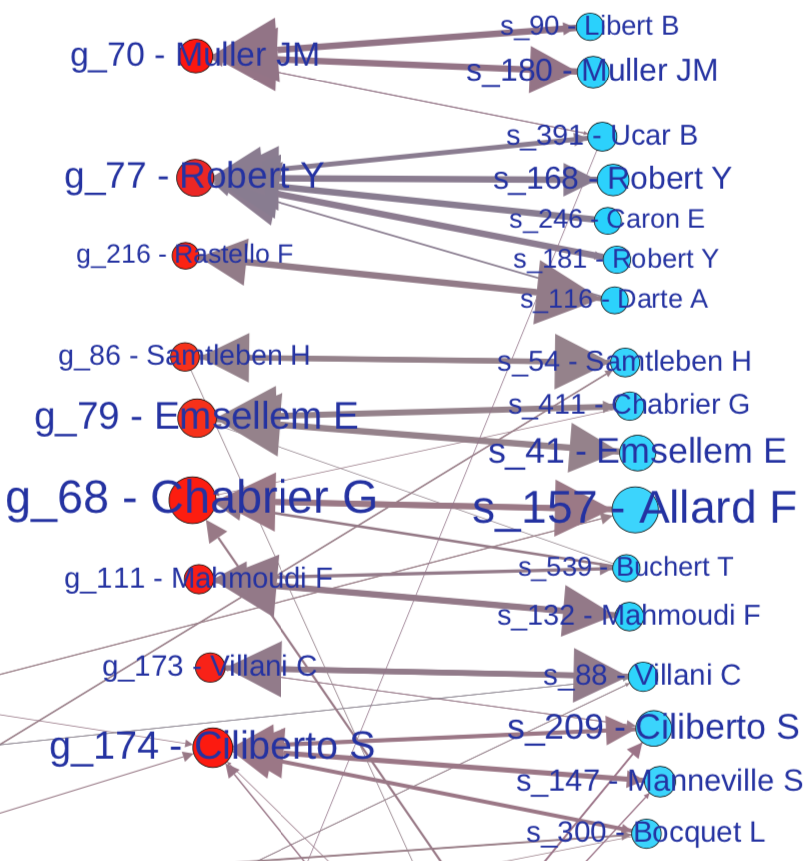}}
\caption[Bipartite network representation of ENS Lyon data set between $P_{GA}$ and $P_{BCLC}$]{Part of the bipartite network representation of ENS Lyon data set. This network shows the links between temporal communities from $P_{GA}$ (left in red) and $P_{BCLC}$ (right in blue). On each node is given the stream ID and the most frequent author name of the temporal community. Size of nodes accounts for the size of the streams, each stream contains at least 20 articles.}
\label{fig:ensl_bipartite_net}
\end{figure}

Figure \ref{fig:ensl_bipartite_net} shows a part of the bipartite network between $P_{GA}$ (left) and $P_{BCLC}$ (right) on the ENS Lyon publications data set. The part of the network is centered on nine streams from $P_{GA}$ equivalent to 17 streams from $P_{BCLC}$. It suggests that streams from $P_{REF}$ are not a mix of different streams from $P_{GA}$ or $P_{BCLC}$. They are rather unions of (almost) entire streams, which means that GA and BCLC yield almost the same partitions, but at different scales. 

\subsection{Results on Wavelets data set}

\subsubsection{Overall comparison}
Describing the history of the wavelets research field is a complicated task as it was born from the collaboration of multiple fields and sub-fields. The values from table \ref{tab:NMIs} show that, even though partitions have a similar number of streams (27 for $P_{GA}$ and 36 for $P_{BCLC}$), there are significant differences between the local and global method. In this case, $NMI(BCLC,REF)$ is significantly higher than $NMI(GA,REF)$ (0.82 vs. 0.68). Moreover $NMI(GA,BCLC)$ is rather low (0.64) which suggests that differences do not only arise from differences of scale. We visualize some of these differences in section \ref{example:wavelets}. From Table \ref{tab:flows} we see that most similar streams between $P_{GA}$ and $P_{REF}$ share $75\% \pm20\%$ of articles on average, whereas the corresponding figure for $P_{BCLC}$ and $P_{REF}$ is $87\% \pm17\%$.

\begin{table}
\centering
\begin{tabular}{rrr}
\hline
Measures & ENS-Lyon & Wavelets \\
\hline
\multirow{2}{*}{$\overline{1^{st}E}(GA,REF)$} & $0.86\pm0.17$ & $0.75\pm0.20$ \\ & $0.49\pm0.20$ & $0.81\pm0.17$ \\\cline{2-3}
\multirow{2}{*}{$\overline{Sum_{80}}(GA,REF)$} & $1.26\pm0.54$ & $1.88\pm0.93$ \\ & $3.37\pm1.76$ & $1.5\pm0.73$ \\\cline{2-3}
\hline
\multirow{2}{*}{$\overline{1^{st}E}(BCLC,REF)$} & $0.89\pm0.14$ & $0.87\pm0.17$ \\ & $0.49\pm0.26$ & $0.87\pm0.15$ \\\cline{2-3}
\multirow{2}{*}{$\overline{Sum_{80}}(BCLC,REF)$} & $1.23\pm0.44$ & $1.26\pm0.50$ \\ & $4.87\pm3.35$ & $1.31\pm0.57$ \\\cline{2-3}
\hline
\multirow{2}{*}{$\overline{1^{st}E}(GA,BCLC)$} & $0.74\pm0.23$ & $0.72\pm0.23$ \\ & $0.85\pm0.16$ & $0.83\pm0.19$ \\\cline{2-3}
\multirow{2}{*}{$\overline{Sum_{80}}(GA,BCLC)$} & $1.96\pm1.14$ & $1.88\pm0.96$ \\ & $1.34\pm0.51$ & $1.61\pm0.83$ \\\cline{2-3}
\hline
\end{tabular}
\caption[Bipartite graph measures]{In this table each cell contains two lines. Each measure $M(X,Y)$ is made on edges. The first line correspond to $M$ measured on edges from $n_X$ to $n_Y$ and the second line corresponds to $M$ being measured on edges from $n_Y$ to $n_X$. So, the first row in $\overline{1^{st}E}(X,Y)$ is the average proportion of articles $n_X$ shares with $n_Y$ $\pm$ its standard deviation. The second row is the average proportion of articles $n_Y$ shares with $n_X$ $\pm$ its standard deviation. For instance, for the ENS-Lyon, this means that streams of $P_{GA}$ share on average 86\% of their articles with their most similar stream in $P_{REF}$, whereas streams from $P_{REF}$ only share on average 49\% of their articles with their most similar stream in $P_{GA}$. $\overline{Sum_{80}}(X,Y)$ is the average number of streams from $P_{Y}$ it takes to retrieve 80\% of the streams' articles from $P_{X}$. For example in the case of the Wavelet data set, on average $1.88\pm 0.96$ streams from $P_{BCLC}$ are needed to retrieve 80\% of a stream from $P_{GA}$.}
\label{tab:flows}
\end{table}

\subsubsection{Examples of major differences}\label{example:wavelets}
We now show some major differences between $P_{GA}$ and $P_{BCLC}$ for the wavelets data set. From Figure \ref{fig:wavelet_bipartite_ga_BCLC}, we can see two types of differences between partitions: scale differences (e.g. g\_9 with s\_51 and s\_166) as in the ENS Lyon case; and more significant differences, when fractions of $P_{GA}$ streams have to be combined to retrieve $P_{BCLC}$ streams (for example, the group of streams around g\_5 and g\_10). Interestingly, g\_7 combines scale and mixing differences. Looking at the BN representation of these $P_{BCLC}$ streams with corresponding $P_{REF}$ streams (Figure \ref{fig:wavelet_bipartite_BCLC_ref}), we see that our $P_{BCLC}$ description is quite similar to the reference description. There are more `stream-to-stream' equivalences, represented by the double arrow on each side of the edge linking streams. Note that, though $P_{BCLC}$ is closer to $P_{REF}$, there are still scale differences (e.g. s\_21, s\_111) and mixing differences (e.g. s\_85, s\_52). 

To understand the origin of the better match of $P_{BCLC}$ to the reference, it is instructive to inspect some of the differences between the local ($P_{BCLC}$) and the global partition $P_{GA}$. Let's look first at the difference between the global method stream g\_7, which corresponds to a merger of four local streams, among which s\_21 and s\_53 (see Fig. \ref{fig:wavelet_bipartite_ga_BCLC}). Figure \ref{fig:streams_wavelets_b} shows that these streams do not belong to the same time period. Stream s\_53 corresponds to the bottom stream (labelled 'Frisch'), and represents early works on wavelets, from 1963 to 1994, focusing on multi fractal analysis and turbulence. The second stream, s\_21 (1987 - 2014, 6th from the top, labelled 'Arneodo'), addressed similar issues in a first period and then, since the early 90's, enlarged the subject matter to include mathematical formalization, together with new applications beyond turbulence, such as genome characterization. Our method appropriately distinguishes these two streams, which correspond to different subfields. The second difference relates to the evolution of one of the authors’ (PF) activities. In the global approach, most PF articles belong to a single cluster that gathers papers in signal representations, and especially time-frequency representations that have been at the heart of his works over the years (third stream starting from the bottom in Fig\ref{fig:streams_wavelets_a}). This is a good approximation, but a finer description of the subjects addressed by PF during his career include three topics: (a) time-frequency methods per se, (b) relations of these methods with wavelets and (c) wavelet methods related to self-similarity, in domains such as turbulence. Using the interactive stream visualization, it is possible to look for $BCLC$ streams containing PF's publications. One finds three streams, addressing the three topics described above, and corresponding to (the stream label refers to its position in Fig\ref{fig:streams_wavelets_b}, starting from the top) respectively streams 3, 8 and 6. These two examples suggest that $BCLC$ is able to capture the complexity of a field dynamics’, including relevant subfields, while the global approach tends to merge streams that represent different fields of inquiry.

\begin{figure}[h]
\centering
\centering
\resizebox*{10.5cm}{!}{\includegraphics{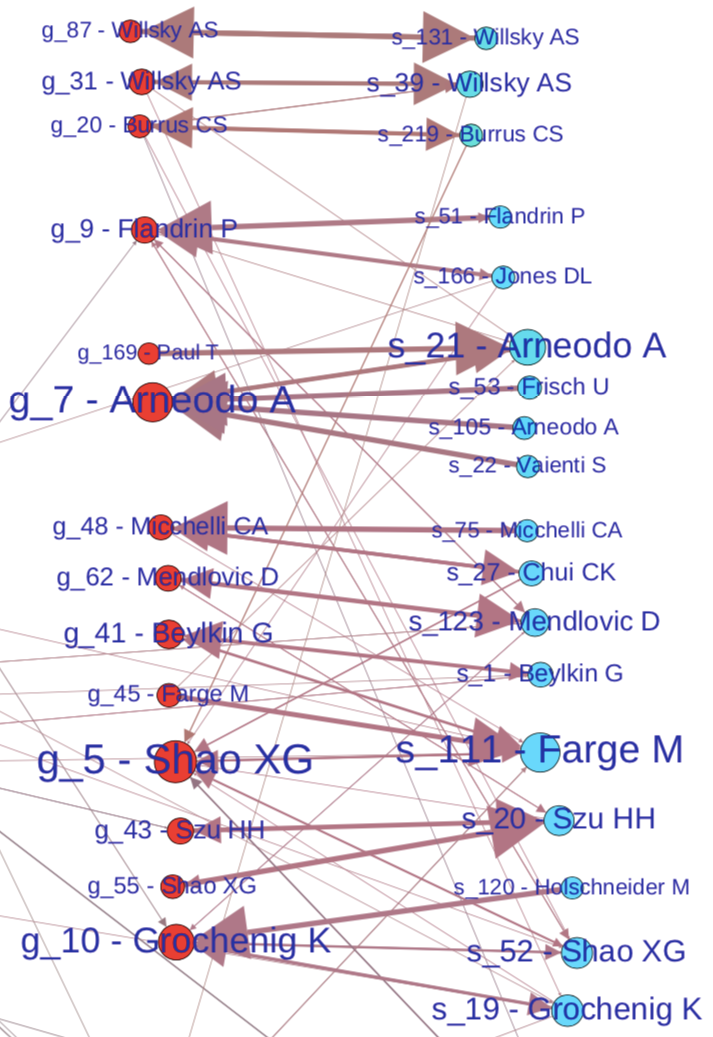}}
\caption[Bipartite network representation of the Wavelets data set ($P_{GA}$ and $P_{BCLC}$)]{Part of the bipartite network representation of Wavelets data set. This network shows the links between temporal communities from $P_{GA}$ (left in red) and $P_{BCLC}$ (right in blue). Each node is labelled by the stream ID and the most frequent author name of the temporal community. Node size accounts for stream size.}
\label{fig:wavelet_bipartite_ga_BCLC}
\end{figure}

\begin{figure}\ContinuedFloat
\centering
\centering
\resizebox*{11cm}{!}{\includegraphics{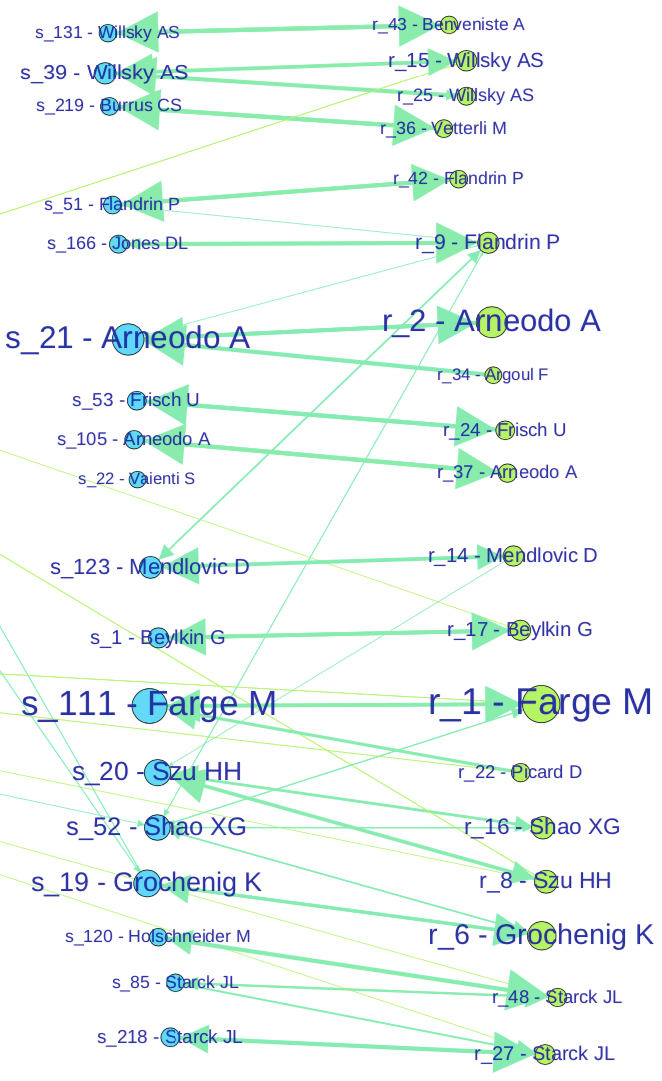}}
\caption[Bipartite network representation of Wavelets data set between $P_{BCLC}$ and $P_{REF}$]{Part of the bipartite network representation of Wavelets data set. This network shows the links between temporal communities from $P_{BCLC}$ (left in blue) and $P_{REF}$ (right in green). Each node is labelled by the stream ID and the most frequent author name of the temporal community. Node size accounts for stream size.}
\label{fig:wavelet_bipartite_BCLC_ref}
\label{fig:wavelet_bipartite}
\end{figure}

\section{Discussion}

We have presented a coherent approach to create a dynamic mesoscopic description of a temporal network. As the standard method to used to create static communities, our method only uses modularity to build the dynamic communities.  We have compared our method to the static (global) approach. We first showed that both methods give the same result for networks with well-separated streams (high modularity), as in the case of ENS-Lyon publications. However, when analyzing data sets with more complex dynamics, as for the birth of wavelets (section \ref{example:wavelets}), our method can generate a more satisfactory dynamics, as compared to an expert reference partition.

Clearly, much more work is needed to develop a standard approach for describing dynamical networks at a mesoscopic scale. The stochastic character of many partitioning algorithms (as Louvain's \cite{blondel_2008}), and the different scales generated by each method make comparisons difficult. Moreover, the dynamical character of the communities renders the definition of an acceptable reference partition even trickier than for static networks.

\newpage
\bibliographystyle{agsm}
\bibliography{refs}

\begin{appendices}

\section{Dynamics of Scientific Research Communities}\label{annex1}
We investigated four temporal community detection methods, two global and two local methods. However, as measures from GA and GPA are very close and measures from BMLA and BCLC are also very close, we only presented the GA and BCLC methods in the core of this article. The two other methods (GPA and BMLA) and their measures are described below.

\subsection{Global Projected Algorithm (GPA)}
Here, we want to include some dynamics into our global algorithm. We thus start with the set of GA-streams obtained by running the Louvain algorithm \cite{blondel_2008} on the global BC network. Then, we define BC networks in each period, only keeping the articles sharing at least two references with at least one other article within the period. Removing the ``long-term connections only'' articles which do not share two or more references with another article in their period results in an average loss of 7.8\% of the articles taken into account in the global BC network. For each time period, we define \emph{local} communities by grouping together the publications that are in the same GA-streams, resulting in a set of local projected communities in each period. Finally, we compute historical streams by applying our matching algorithm to the projected communities. Interestingly, the streams that are build from this method do not necessarily correspond to the GA-streams: the predecessors / successors of a cluster may not be subsets of the same GA-stream of this particular cluster, resulting in splits or merges. In practice, a few GA-streams may in effect be cut into into two or more GPA-streams localized in different time periods. This approach thus allows to visualize the evolution of a GA-stream in terms of dynamical events (splits and merges).\\

\subsection{Best-Modularity Local Algorithm (BMLA)}
For each time period, we run $N$ independent runs (we used $N=100$) of the Louvain algorithm. Because of the noise inherent to the Louvain algorithm, these partitions may be a bit different, while having similar modularity values (in practice the modularity difference between the partitions of different runs is lower than 0.005). Compared to the BCLC method, we do not try here to choose the partitions of the run best matching the partition from the previous or next period, but keep the partition with the best modularity among the $N$ runs in each time period. BMLA historical streams are then defined by applying the matching algorithm to these `best-modularity' partitions.

{\bf BMLA Algorithm}\\
\begin{algorithm}[H]
 Compute the Bibliographic Coupling Graph \;
 Split the data set into temporal windows $\Delta t$ \;
 \For{each temporal window}{
  run $N=100$ Louvain algorithm on the instant network\;
  select the instant partition with the highest modularity $Q$\;
  }
 Match the most similar communities between successive temporal windows \;
 Link the paired communities along time\;
 \ 
\end{algorithm}

This algorithm returns temporal streams we call \emph{BMLA-streams}. These streams maximize the modularity at each time $t$ without considering the global modularity of the whole system.

\subsection{Comparing All Algorithms}

\begin{table}
\centering
\begin{tabular}{rrr}
\hline
Measures & ENS-Lyon & Wavelets \\
\hline
$\vert P_{GA}\vert$ & 57 & 27 \\
$\vert P_{GPA}\vert$ & 54 & 30 \\
$\vert P_{BCLC}\vert$ & 97 & 36 \\
$\vert P_{BMLA}\vert$ & 103 & 40 \\
$\vert P_{REF}\vert$ & 17 & 36 \\
\hline
$H(P_{GA})$ & 3.63 & 2.87 \\
$H(P_{GPA})$ & 3.63 & 2.94 \\
$H(P_{BCLC})$ & 4.05 & 3.04 \\
$H(P_{BMLA})$ & 4.04 & 3.17 \\
$H(P_{REF})$ & 2.37 & 3.18 \\
\hline
$MI(GA,REF)$ & 1.93 & 2.03 \\
$MI(GPA,REF)$ & 1.94 & 2.09 \\
$MI(BCLC,REF)$ & 1.93 & 2.49 \\
$MI(BMLA,REF)$ & 1.94 & 2.47 \\
$MI(GA,BCLC)$ & 3.10 & 1.90 \\
\hline
$NMI_{GA}(GA,REF)$ & 0.53 & 0.73 \\
$NMI_{REF}(GA,REF)$ & 0.82 & 0.64 \\
$NMI(GA,REF)$ & 0.66 & 0.68 \\
\hline
$NMI_{GPA}(GPA,REF)$ & 0.54 & 0.74 \\
$NMI_{REF}(GPA,REF)$ & 0.82 & 0.66 \\
$NMI(GPA,REF)$ & 0.67 & 0.70 \\
\hline
$NMI_{BCLC}(BCLC,REF)$ & 0.48 & 0.84 \\
$NMI_{REF}(BCLC,REF)$ & 0.81 & 0.80 \\
$NMI(BCLC,REF)$ & 0.63 & 0.82 \\
\hline
$NMI_{BMLA}(BMLA,REF)$ & 0.48 & 0.78 \\
$NMI_{REF}(BMLA,REF)$ & 0.82 & 0.80 \\
$NMI(BMLA,REF)$ & 0.63 & 0.79 \\
\hline
$NMI_{GA}(GA,BCLC)$ & 0.86 & 0.67 \\
$NMI_{BCLC}(GA,BCLC)$ & 0.77 & 0.62 \\
$NMI(GA,BCLC)$ & 0.81 & 0.64 \\
\hline
\end{tabular}
\caption[Entropies and Mutual Information measures for all algorithms]{Similarly to Table \ref{tab:NMIs}, $\vert P_{X}\vert$ is the number of streams in partition $X$. $H(P_{X})$ is the entropy of partition $X$. $MI(P_{X},P_{Y})$ is the mutual information between the partitions $X$ and $Y$. $NMI_{X}$ is the mutual information MI normalized by $H(P_{X})$. $NMI(P_{X},P_{Y})$ is the symmetrical normalized mutual information (normalized by $\sqrt{H(X)*H(Y)}$).}
\label{tab:NMIs_complete}
\end{table}

\begin{table}
\centering
\begin{tabular}{rrr}
\hline
Measures & ENS-Lyon & Wavelets \\
\hline
\multirow{2}{*}{$\overline{1^{st}E}(GA,REF)$} & $0.86\pm0.17$ & $0.75\pm0.20$ \\ & $0.49\pm0.20$ & $0.81\pm0.17$ \\\cline{2-3}
\multirow{2}{*}{$\overline{Sum_{80}}(GA,REF)$} & $1.26\pm0.54$ & $1.88\pm0.93$ \\ & $3.37\pm1.76$ & $1.5\pm0.73$ \\\cline{2-3}
\hline
\multirow{2}{*}{$\overline{1^{st}E}(GPA,REF)$} & $0.87\pm0.16$ & $0.78\pm0.19$ \\ & $0.54\pm0.23$ & $0.83\pm0.17$ \\\cline{2-3}
\multirow{2}{*}{$\overline{Sum_{80}}(GPA,REF)$} & $1.24\pm0.5$ & $1.65\pm0.84$ \\ & $3.12\pm1.61$ & $1.47\pm0.72$ \\\cline{2-3}
\hline
\multirow{2}{*}{$\overline{1^{st}E}(BCLC,REF)$} & $0.89\pm0.14$ & $0.87\pm0.17$ \\ & $0.49\pm0.26$ & $0.87\pm0.15$ \\\cline{2-3}
\multirow{2}{*}{$\overline{Sum_{80}}(BCLC,REF)$} & $1.23\pm0.44$ & $1.26\pm0.50$ \\ & $4.87\pm3.35$ & $1.31\pm0.57$ \\\cline{2-3}
\hline
\multirow{2}{*}{$\overline{1^{st}E}(BMLA,REF)$} & $0.89\pm0.14$ & $0.85\pm0.19$ \\ & $0.49\pm0.25$ & $0.84\pm0.17$ \\\cline{2-3}
\multirow{2}{*}{$\overline{Sum_{80}}(BMLA,REF)$} & $1.23\pm0.44$ & $1.34\pm0.63$ \\ & $5.0\pm3.60$ & $1.37\pm0.59$ \\\cline{2-3}
\hline
\multirow{2}{*}{$\overline{1^{st}E}(GA,BCLC)$} & $0.74\pm0.23$ & $0.72\pm0.23$ \\ & $0.85\pm0.16$ & $0.83\pm0.19$ \\\cline{2-3}
\multirow{2}{*}{$\overline{Sum_{80}}(GA,BCLC)$} & $1.96\pm1.14$ & $1.88\pm0.96$ \\ & $1.34\pm0.51$ & $1.61\pm0.83$ \\\cline{2-3}
\hline
\end{tabular}
\caption[Bipartite graph measures for all algorithms]{Similarly to Table \ref{tab:flows}, In this table each cell contains two lines. Each measure $M(X,Y)$ is made on edges. The first line correspond to $M$ measured on edges from $n_X$ to $n_Y$ and the second line corresponds to $M$ being measured on edges from $n_Y$ to $n_X$. So, the first row in $\overline{1^{st}E}(X,Y)$ is the average proportion of articles $n_X$ shares with $n_Y$ $\pm$ its standard deviation. The second row is the average proportion of articles $n_Y$ shares with $n_X$ $\pm$ its standard deviation. For instance, for the ENS-Lyon, this means that streams of $P_{GA}$ share on average 86\% of their articles with their most similar stream in $P_{REF}$, whereas streams from $P_{REF}$ only share on average 49\% of their articles with their most similar stream in $P_{GA}$. $\overline{Sum_{80}}(X,Y)$ is the average number of streams from $P_{Y}$ it takes to retrieve 80\% of the streams' articles from $P_{X}$. For example in the case of the Wavelet data set, on average $1.88\pm 0.96$ streams from $P_{BCLC}$ are needed to retrieve 80\% of a stream from $P_{GA}$.}
\label{tab:flows_complete}
\end{table}

Table \ref{tab:NMIs_complete} and Table \ref{tab:flows_complete} show there is very little difference between the local algorithms and between the global algorithms, for all measures on both data sets.

\end{appendices}

\end{document}